\documentclass[12pt,a4paper]{article}
\usepackage{a4wide}
\usepackage{latexsym}
\usepackage{epsf}
\usepackage{amssymb}

\begin{document}

\begin{flushright}
\small
IFT-UAM/CSIC-02-23\\
{\bf hep-th/0206159}\\
June $17$th, $2002$
\normalsize
\end{flushright}

\begin{center}


\vspace{2cm}

{\Large {\bf A Note on Lie-Lorentz Derivatives}}

\vspace{2cm}


{\bf\large Tom\'as Ort\'{\i}n} \\

\vspace{.5cm}
 {\tt Tomas.Ortin@cern.ch}
\vskip 1truecm

\vskip 0.2cm 
{\it Instituto de F\'{\i}sica Te\'orica, C-XVI,
Universidad Aut\'onoma de Madrid \\
E-28049-Madrid, Spain}

{\it and}

{\it I.M.A.F.F., C.S.I.C., 
Calle de Serrano 113 bis\\ 
E-28006-Madrid, Spain}

\vspace{2cm}


{\bf Abstract}

\end{center}

\begin{quotation}

\small

The definition of ``Lie derivative'' of spinors with respect to
Killing vectors is extended to all kinds of Lorentz tensors. This
Lie-Lorentz derivative appears naturally in the commutator of two
supersymmetry transformations generated by Killing spinors and
vanishes for Vielbeins. It can be identified as the generator of the
action of isometries on supergravity fields and its use for the
calculation of supersymmetry algebras is revised and extended.

\end{quotation}

\newpage

\pagestyle{plain}


\section*{Introduction}

Spinors are defined by their transformation properties under
$SO(n_{+},n_{-})$ (``Lorentz'') transformations and, thus, they can only
be introduced in the tangent space of curved spaces using the formalism
introduced by Weyl in Ref.~\cite{kn:Weyl4}. This formalism makes use of
Vielbeins $\{e_{a}{}^{\mu}\}$ which form an orthonormal basis in tangent
space

\begin{equation}
e_{a}{}^{\mu}e_{b}{}^{\nu} g_{\mu\nu}=\eta_{ab}\, ,
\hspace{1cm}
(\eta_{ab})={\rm diag}\,(+\cdots+,-\cdots -)\, ,
\end{equation}

\noindent and it is covariant w.r.t.~local transformations that preserve
this orthonormality: local $SO(n_{+},n_{-})$ transformations. These
are the only transformations that act non-trivially on any Lorentz
tensor\footnote{We will call Lorentz tensor any object transforming in
  some finite-dimensional representation of the Lorentz group like
  vectors and spinors.}  $T$ in the representation $r$

\begin{equation}
T^{\prime}(x^{\prime}) = \Gamma_{r}[g(x)] T(x)\, ,
\end{equation}

\noindent where $\Gamma_{r}[g(x)]$ is the representation of the
position-dependent group element $g(x)$ that can be constructed by
exponentiation

\begin{equation}
\Gamma_{r}[g(x)]=
{\rm exp}[{\textstyle\frac{1}{2}} \sigma^{ab}(x)
\Gamma_{r}\left(M_{ab}\right)] 
\end{equation}

\noindent where $\Gamma_{r}\left(M_{ab}\right)$ are the
$SO(n_{+},n_{-})$ generators in the representation $r$. For the
contravariant vector and spinor representations

\begin{equation}
\Gamma_{v}\left(M_{ab}\right){}^{c}{}_{d}
=2\eta_{[a}{}^{c}\eta_{b]d}\, ,
\hspace{1cm}
\Gamma_{s}\left(M_{ab}\right) 
={\textstyle\frac{1}{2}}\Gamma_{[a}\Gamma_{b]}\, ,
\hspace{1cm}
\{\Gamma_{a},\Gamma_{b}\}=2\eta_{ab}\, .
\end{equation}

Local Lorentz covariance is required in order to be able to gauge away
the additional degrees of freedom that the Vielbein ($d^{2}$) has, as
compared with the metric ($d(d+1)/2$), and it is achieved by
introducing a covariant derivative $\mathcal{D}_{\mu}$ which acts on
$T$ according to

\begin{equation}
\mathcal{D}_{\mu}T\equiv \left(\partial_{\mu}
-\omega_{r\, \mu}\right) T\, ,
\hspace{1cm}
\omega_{r\, \mu} \equiv
{\textstyle\frac{1}{2}}\omega_{\mu}{}^{ab}
\Gamma_{r}\left(M_{ab}\right) \, .
\end{equation}

\noindent $\omega_{\mu}{}^{ab}$ is the $SO(n_{+},n_{-})$ (spin) 
connection, i.e.~it is antisymmetric $\omega_{\mu}{}^{ab}=
  -\omega_{\mu}{}^{ba}$, which implies 

\begin{equation}
\label{eq:constantmetric}
\mathcal{D}_{\mu}\eta_{ab}=0\, .
\end{equation}

\noindent and transforms according to

\begin{equation}
\omega_{r\, \mu}^{\prime}=\Gamma_{r}[g(x)] 
\omega_{r\, \mu} \Gamma^{-1}_{r}[g(x)] 
+(\partial_{\mu}\Gamma_{r}[g(x)] )\Gamma^{-1}_{r}[g(x)]\, .
\end{equation}

In order to have only one connection the spin connection is related to
the affine connection $\Gamma_{\mu\nu}{}^{\rho}$ by the Vielbein
postulate

\begin{equation}
\nabla_{\mu} e_{a}{}^{\nu} = \partial_{\mu} e_{a}{}^{\nu} 
+\Gamma_{\mu\rho}{}^{\nu}e_{a}{}^{\rho} 
-e_{b}{}^{\nu}\omega_{\mu a}{}^{b}=0\, ,
\end{equation}

\noindent ($\nabla_{\mu}$ denotes the total (general and Lorentz)
covariant derivative) which implies

\begin{equation}
\label{eq:relationbetweentheconnections}
\omega_{\mu\, a}{}^{b} = \Gamma_{\mu a}{}^{b} 
-e_{a}{}^{\nu}\partial_{\mu}e_{\nu}{}^{b}\, .  
\end{equation}

\noindent The Vielbein postulate together with 
Eq.~(\ref{eq:constantmetric}) imply that the affine connection is
metric compatible and the spin connection is completely determined by
the Vielbein, their inverses, and the contorsion tensor $K_{abc}$:

\begin{equation}
\omega_{abc}=-\Omega_{abc} +\Omega_{bca} -\Omega_{cab}+K_{abc}\, ,
\hspace{1cm}
\Omega_{abc} =e_{a}{}^{\mu} e_{b}{}^{\nu}\partial_{[\mu|}e_{c |\nu]}\, .
\end{equation}


In Weyl's formalism Lorentz tensors (in particular spinors) behave,
then, as scalars under general coordinate transformations (g.c.t.'s).
However, we can use Weyl's formalism just to work in curvilinear
coordinates in Minkowski spacetime and we would find that a standard
Lorentz transformation becomes a g.c.t.  that does not act on the
spinorial indices. This looks strange, but it is unavoidable in Weyl's
formalism.

On the other hand, Lorentz transformations and g.c.t.'s do not commute
and the result of a g.c.t.~on a Lorentz tensor is strongly
frame-dependent. This shows up in the standard Lie derivative (an
infinitesimal g.c.t.) which does not transform Lorentz tensors into
Lorentz tensors. Thus, while the Lie derivative of the metric
$g_{\mu\nu}$ with respect to a Killing vector field vanishes (by
definition) the Lie derivative of the inverse Vielbein $e^{a}{}_{\mu}$
associated to the same metric with respect to the same Killing vector
in general does not.

For spinors $\psi$ this situation was solved in
Ref.~\cite{kn:Kos}\footnote{For a different approach an further
  references, see, e.g.~\cite{Hurley:cf}. The Lie derivative on
  spinors has been studied and used in the derivation of supersymmetry
  algebras in Refs.~\cite{Vandyck:ei,Vandyck:gc} and more recently in
  Ref.~\cite{Figueroa-O'Farrill:1999va} whose results to revise and
  extend.} where a {\it spinorial Lie derivative} $\mathbb{L}_{k}$
with respect to Killing vector fields $k$ was defined by\footnote{We
  use the symbol $\mathbb{L}_{k}$ to avoid confusion with the standard
  Lie derivative that we keep denoting by $\mathcal{L}_{k}$. On
  spinors, then, $\mathcal{L}_{k}\psi =k^{\mu}\partial_{\mu}\psi$.}

\begin{equation}
\label{eq:LLonspinors}
\mathbb{L}_{k}\psi \equiv k^{\mu}\mathcal{D}_{\mu}\psi 
+{\textstyle\frac{1}{4}} \mathcal{D}_{[a}k_{b]}\Gamma^{ab}\psi\, .  
\end{equation}

This derivative is a derivation that transforms spinors into spinors
(it is Lorentz-covariant) and satisfies the property 

  \begin{equation}
   [\mathbb{L}_{k_{1}},\mathbb{L}_{k_{2}}]\, \psi 
   =\mathbb{L}_{[k_{1},k_{2}]}\, \psi\, ,
  \end{equation}

\noindent but, most importantly, the second term 
defines an action of certain g.c.t.'s (the isometries) on the
spinorial indices: an infinitesimal Lorentz transformation with
parameter $\frac{1}{2}\mathcal{D}_{[a}k_{b]}$. This is precisely what
one expects on physical grounds. Let us consider an example: Minkowski
spacetime $g_{\mu\nu}=\eta_{\mu\nu}$ with the obvious Vielbein
$e_{a}{}^{\mu}=\delta_{a}{}^{\mu}$ and the  infinitesimal
g.c.t.~generated by the Killing vector

\begin{equation}
k^{\mu} = \sigma^{\mu}{}_{\nu}x^{\nu}\, ,  
\end{equation}

\noindent where  $\sigma^{\mu\nu}=-\sigma^{\nu\mu}$  and constant. 
This is a standard infinitesimal Lorentz transformation and, indeed,
we find

\begin{equation}
\mathbb{L}_{k}=k^{\mu}\partial_{\mu}\psi 
+{\textstyle\frac{1}{4}}\sigma_{ab}\Gamma^{ab}\psi\, ,   
\end{equation}

\noindent which is the result that one would obtain in 
the standard spinor formalism.

The spinorial Lie derivative can be seen as a Lorentz covariantization
of the standard Lie derivative (first term) supplemented by an
infinitesimal local Lorentz transformation that trivializes the
holonomy. It is clear that these ideas can be extended to other
Lorentz tensors and we can define a Lie-lorentz derivative which is
first a Lorentz covariantization of the standard Lie derivative
supplemented by an infinitesimal Lorentz transformation that
trivializes the holonomy.  The parameter of this transformation has to
be exactly the same as in the spinorial case and, thus, we arrive to
the following definition.


\section{Definition and Properties of the Lie-Lorentz Derivative}

On pure Lorentz tensors $T$ we define the Lie-Lorentz derivative with
respect to the Killing vector $k$ by

\begin{equation}
\mathbb{L}_{k}T \equiv k^{\rho}\nabla_{\rho}T +{\textstyle\frac{1}{2}}
\nabla_{[a}k_{b]}\, \Gamma_{r}(M^{ab})T\, .    
\end{equation}

\noindent On tensors  that also have world indices
$T_{\mu_{1}\cdots\mu_{m}}{}^{\nu_{1}\cdots\nu_{n}}$ 

\begin{equation}
  \begin{array}{rcl}
\mathbb{L}_{k}T_{\mu_{1}\cdots\mu_{m}}{}^{\nu_{1}\cdots\nu_{n}}
& \equiv & 
k^{\rho}\nabla_{\rho}T_{\mu_{1}\cdots\mu_{m}}{}^{\nu_{1}\cdots\nu_{n}} 
-\nabla_{\rho}k^{\nu_{1}}
T_{\mu_{1}\cdots\mu_{m}}{}^{\rho\nu_{2}\cdots\nu_{n}}
-\cdots \\
& & \\
& & 
+\nabla_{\mu_{1}}k^{\rho}
T_{\rho\mu_{2}\cdots\mu_{m}}{}^{\nu_{1}\cdots\nu_{n}}
+\cdots
+{\textstyle\frac{1}{2}}
\nabla_{[a}k_{b]}\, \Gamma_{r}(M^{ab})
T_{\mu_{1}\cdots\mu_{m}}{}^{\nu_{1}\cdots\nu_{n}}\, .    \\
\end{array}
\end{equation}

\noindent In all the cases that  we are going to consider 
$\nabla_{\mu}$ is the full (affine plus Lorentz) torsionless covariant
derivative satisfying the Vielbein postulate.

In the following $T_{1},T_{2}$ will be two mixed tensors of any kind,
$k_{1},k_{2}$ any two conformal Killing vector fields and
$a^{1},a^{2}$ two arbitrary constants. The Lie-Lorentz derivative has
the following basic properties:

\begin{enumerate}

\item  Leibnitz rule:

  \begin{equation}
  \label{eq:Leibnitz}
  \mathbb{L}_{k}(T_{1}T_{2})= \mathbb{L}_{k}(T_{1})T_{2} 
 +T_{1} \mathbb{L}_{k}T_{2}\, .
  \end{equation}

 Thus, it is a derivation.

\item The commutator of two  Lie-Lorentz derivatives
  
  \begin{equation}
\label{eq:bracket}
   [\mathbb{L}_{k_{1}},\mathbb{L}_{k_{2}}]\, T 
   =\mathbb{L}_{[k_{1},k_{2}]}\, T\, ,
  \end{equation}
  
  where $[k_{1},k_{2}]$ is their Lie bracket.
  
\item The Lie-Lorentz derivative is linear in the vector fields

  \begin{equation}
   \mathbb{L}_{a^{1}k_{1}+a^{2}k_{2}}\, T =  a^{1}\, \mathbb{L}_{k_{1}} T
 +a^{2}\, \mathbb{L}_{k_{2}}T\, ,
  \end{equation}
  
  and, thus, the Lie-Lorentz derivative with respect to the conformal
  killing vector fields forms a representation of the Lie algebra of
  conformal isometries of the manifold.

\end{enumerate}

Some immediate consequences of the definition and basic properties
are:

\begin{enumerate}

\item The Lie-Lorentz derivative of the Vielbein is 

  \begin{equation}
   \label{eq:LLvielbein}
    \mathbb{L}_{k}e^{a}{}_{\mu}
    ={\textstyle\frac{1}{d}}\nabla_{\rho}k^{\rho}e^{a}{}_{\mu}\, ,
  \end{equation}
  
  and  vanishes when $k$ is a Killing vector. In this case, we have
  the desirable property

  \begin{equation}
    \mathbb{L}_{k}\xi^{a} = e^{a}{}_{\mu} \mathcal{L}_{k}\xi^{\mu}\, .
  \end{equation}

\item The Lie-Lorentz derivative of gamma matrices is zero.

  \begin{equation}
   \label{eq:LLgamma}
    \mathbb{L}_{k}\gamma^{a}=0\, .
  \end{equation}
  
\item \label{item:1} As a consequence of Eqs.~(\ref{eq:Leibnitz}),
  (\ref{eq:LLvielbein}) and (\ref{eq:LLgamma}), the Lie-Lorentz
  derivative with respect to Killing vectors preserves the Clifford
  action of vectors $v$ on spinors $\psi$ $v\cdot \psi \equiv
  v_{a}\Gamma^{a} \psi=\not\!v \psi$ :

  \begin{equation}
  \label{eq:preserve1}
   [\mathbb{L}_{k},\not\!v]\, \psi = [k,v]\cdot \psi\, . 
  \end{equation}

\item \label{item:2} Also for Killing vectors $k$ it preserves the covariant
  derivative

  \begin{equation}
  \label{eq:preserve2}
   [\mathbb{L}_{k},\nabla_{v}]\, T = \nabla_{[k,v]}\, T\, . 
  \end{equation}
  
\item \label{item:3} All these properties imply that the Lie-Lorentz
  derivative with respect to Killing vectors preserves the
  supercovariant derivative of supergravity theories, at least in the
  simplest cases that we are going to examine next.

\end{enumerate}

We stress that the properties \ref{item:1}, \ref{item:2}, \ref{item:3}
are only valid for Killing (not just conformal Killing) vectors.


\section{The Lie-Lorentz Derivative and Supersymmetry}

The Lie-Lorentz derivative occurs naturally in supergravity theories.

To start with, let us consider the local on-shell supersymmetry
algebra, in particular the commutator of two infinitesimal, local
supersymmetry transformations $\delta_{Q}(\epsilon)$ in $N=1,d=4$
supergravity 

\begin{equation}
\delta_{Q}(\epsilon)\, e^{a}{}_{\mu} = 
-i\bar{\epsilon}\gamma\psi_{\mu}\, ,
\hspace{1cm}
\delta_{Q}(\epsilon)\, \psi_{\mu}  = 
{\cal D}_{\mu}\epsilon\, .
\end{equation}

\noindent which is usually written in the form

\begin{equation}
\label{eq:commutator1}
[\delta_{Q}(\epsilon_{1}), \delta_{Q}(\epsilon_{2})] =\delta_{\rm gct}(\xi)  
+\delta_{\rm LL}(\sigma) +\delta_{Q} (\epsilon)\, ,
\end{equation}

\noindent where $\delta_{\rm gct}(\xi)$  is an infinitesimal general 
coordinate transformation with parameter 

\begin{equation}
\xi^{\mu} = -i\bar{\epsilon}_{1}\gamma^{\mu}\epsilon_{2}\, .
\end{equation}

\noindent and is given by $\delta_{\rm gct}(\xi)=-\pounds_{\xi}$, 
$\delta_{\rm LL}(\sigma)$ is an infinitesimal local Lorentz
transformation with parameter 

\begin{equation}
\sigma^{ab} = \xi^{\nu}\omega_{\nu}{}^{ab}\, ,
\end{equation}

\noindent and where 

\begin{equation}
\epsilon = \xi^{\mu}\psi_{\mu}\, .
\end{equation}

We are interested in obtaining the global superalgebra from the
commutators of all the symmetry transformations of the theory. There
is no unique superalgebra associated to a given supergravity theory,
rather there are superalgebras associated to given bosonic solutions
of the supergravity theory and their (super) symmetries. Solutions
with a high degree of (super) symmetry are usually considered vacua of
the supergravity theory and their associated superalgebras are of
special interest.

Let us, then, consider a given vacuum solution of the $N=1,d=4$
supergravity equations of motion admitting Killing spinors
$\varepsilon$ and Killing vectors $k$

\begin{equation}
\mathcal{D}_{\mu}  \varepsilon =0\, ,
\hspace{1cm}
\nabla_{(\mu}k_{\nu)}=0\, .
\end{equation}

\noindent  On this vacuum, the commutator
Eq.~(\ref{eq:commutator1}) should reduce to the commutator of two
supercharges $Q_{1,2}=\bar{\varepsilon}_{1,2}Q$ which should be
proportional to translations $\sim k^{a}P_{a}$ where $k^{a}=-i
\bar{\varepsilon}_{1}\gamma^{a}\varepsilon_{2}$ is a Killing vector
if, as we have assumed, $\varepsilon_{1,2}$ are Killing spinors.
However, the commutator Eq.~(\ref{eq:commutator1}) does not give that
result if we naively interpret $\delta_{\rm gct}(k)$ as an
infinitesimal translation since there is another term $\delta_{\rm
  LL}$ whose meaning is unclear.

Observing that, actually, all the Killing vectors $k^{a}=-i
\bar{\varepsilon}_{1}\gamma^{a}\varepsilon_{2}$ are covariantly constant,
we can write instead 

\begin{equation}
[\delta_{Q}(\varepsilon_{1}), \delta_{Q}(\varepsilon_{2})]=
\delta_{P}(k)\, ,
\hspace{1cm}
k^{a}=-i \bar{\varepsilon}_{1}\gamma^{a}\varepsilon_{2}
\end{equation}

\noindent where now we have identified the generator of the vacuum 
isometries acting on the supergravity fields

\begin{equation}
\delta_{P}(k) \equiv -\mathbb{L}_{k}\, .
\end{equation}

This commutator has an immediate interpretation in terms of the global
superalgebra. On the other hand, in this form, on account of
Eq.~(\ref{eq:LLvielbein}), it is evident the the commutator of two
supersymmetry transformations generated by two Killing spinors leaves
invariant the Vierbein, as it should be.

To check that these definitions and interpretations actually make
sense, let us consider a more complicated case: gauged $N=2,d=4$
supergravity, whose supersymmetry transformation rules are

\begin{equation}
\delta_{Q}(\epsilon)\,  e^{a}{}_{\mu} =
-i\bar{\epsilon}\gamma^{a}\psi_{\mu}\, ,
\hspace{1cm}
\delta_{Q}(\epsilon)\,  A_{\mu} =
-i\bar{\epsilon}\sigma^{2}\psi_{\mu}\, ,
\hspace{1cm}
\delta_{Q}(\epsilon)\,  \psi_{\mu} =
\tilde{\mathcal{D}}_{\mu}\epsilon\, ,
\end{equation}

\noindent where

\begin{equation}
\label{eq:supercovariantderivativeungauged}
\tilde{\nabla}_{\mu}= \hat{\mathcal{D}}_{\mu} +ig A_{\mu}\sigma^{2}
+{\textstyle\frac{1}{4}}\not\!\tilde{F}\gamma_{\mu}\sigma^{2}\, ,
\end{equation}

\noindent is the {\it supercovariant derivative}
and 

\begin{equation}
\hat{\mathcal{D}}_{\mu}  =
\mathcal{D}_{\mu} -{\textstyle\frac{ig}{2}}\gamma_{\mu}\, , 
\end{equation}

\noindent is the $AdS_{4}\sim SO(2,3)$ covariant derivative. 
The commutator of two $N=2,d=4$ supersymmetry transformations is
usually written in this form

\begin{equation}
\label{eq:commutator2}
[\delta_{Q}(\epsilon_{1}), \delta_{Q}(\epsilon_{2})] =\delta_{\rm gct}(\xi)  
+\delta_{\rm LL}(\sigma) +\delta_{e}(\Lambda) +\delta_{Q} (\epsilon)\, ,
\end{equation}

\noindent where the parameter of the infinitesimal local Lorentz
transformations is now

\begin{equation}
\sigma^{ab}  =  
\xi^{\mu}\omega_{\mu}{}^{ab} -g\bar{\epsilon}_{2}\gamma^{ab}\epsilon_{1}
-i \bar{\epsilon}_{2}\left(\tilde{F}^{ab}
-i\gamma_{5}{}^{\star}\tilde{F}^{ab}\right) 
\sigma^{2}\epsilon_{1}\, ,
\end{equation}

\noindent and where $\delta_{e}(\Lambda)$ are $U(1)$ gauge transformations
of the vector field and charged gravitino with parameter

\begin{equation}
\Lambda =  -i\bar{\epsilon}_{2}\sigma^{2}\epsilon_{1}
+\xi^{\nu}A_{\nu}\, .
\end{equation}

It is easy to see that on a vacuum solution admitting Killing spinors
$\varepsilon$ and Killing vectors $k$ 

\begin{equation}
\tilde{\nabla}_{\mu}  \varepsilon =0\, ,
\hspace{1cm}
\nabla_{(\mu}k_{\nu)}=0\, .
\end{equation}

\noindent the commutator Eq.~(\ref{eq:commutator2}) 
can be written in the form

\begin{equation}
[\delta_{Q}(\varepsilon_{1}), \delta_{Q}(\varepsilon_{2})]=
\delta_{P}(k) +\delta_{e}(\chi)\, ,
\hspace{1cm}
k^{a}=-i \bar{\varepsilon}_{1}\gamma^{a}\varepsilon_{2}\, ,
\hspace{.5cm}
\chi =  -i\bar{\varepsilon}_{2}\sigma^{2}\varepsilon_{1}
+k^{\nu}A_{\nu}\, .
\end{equation}

The interpretation is again immediate.  This commutator should vanish
on all the fields of the theory. It clearly does on the Vierbein. On
the vector and gravitino fields, it tells us that the Lie derivative
vanishes up to a gauge transformation.


Let us consider now the remaining commutators. 

Since we are identifying the bosonic generators of the the bosonic
subalgebra of the supersymmetry algebra of the vacuum with the
Lie-Lorentz derivative with respect to the Killing vector fields of
the solution, we are going to get as bosonic subalgebra the Lie
algebra of isometries of the vacuum, on account of
Eq.~(\ref{eq:bracket}):

\begin{equation}
[\delta_{P}(k_{1}),\delta_{P}(k_{2})]
=\delta_{P}([k_{1},k_{2}])\, .
\end{equation}

The commutator of local supersymmetry transformations and g.c.t.'s.
these are difficult to compute, as a matter of principle, since the
standard Lie derivative of Lorentz tensors is not a Lorentz tensor and
then it does not make sense to perform on the transformed tensor a
further supersymmetry transformation. Further, while we have a
prescription to calculate the effect of an infinitesimal g.c.t.~on any
geometrical object, we do not know how to calculate the effect of an
infinitesimal supersymmetry transformation of geometrical objects which
are not fields of our theory (for instance, the vector that generates
the infinitesimal g.c.t.).

Thus, it is necessary to give a prescription to calculate these
commutators. In Ref.~\cite{Figueroa-O'Farrill:1999va} the following
prescription was proposed:

\begin{equation}
[\delta_{Q}(\varepsilon),\delta_{P}(k)]
=\delta_{Q}(\mathbb{L}_{k}\varepsilon)\, .
\end{equation}

This prescription is based on the property (checked for certain
geometrical Killing spinors in that reference) that the Lie-Lorentz
derivative preserves the supercovariant derivative and, therefore,
transforms Killing spinors into Killing spinors. 

This property also holds in the simple theories we are considering
here: on account of Eqs.~(\ref{eq:preserve1}),(\ref{eq:preserve2}), we
get

\begin{equation}
[\mathbb{L}_{k},\hat{\mathcal{D}}_{v}] \epsilon 
=\hat{\mathcal{D}}_{[k,v]} \epsilon\, , 
\end{equation}

\noindent (so the $AdS_{4}$ covariant derivative is preserved 
and the Lie-Lorentz derivative transforms $N=1,AdS_{4}$ Killing
spinors into Killing spinors \cite{Figueroa-O'Farrill:1999va}) and, if

\begin{equation}
\label{eq:thecondition}
\mathbb{L}_{k}A_{a}=0\, ,  
\end{equation}

\noindent we find that the $N=2,AdS_{4}$ supercovariant derivative
is also preserved 

\begin{equation}
[\mathbb{L}_{k},\tilde{\mathcal{D}}_{v}] \epsilon 
=\tilde{\mathcal{D}}_{[k,v]} \epsilon\, , 
\end{equation}

\noindent and the Lie-Lorentz derivative transforms again Killing
spinors into Killing spinors.

The condition Eq.~(\ref{eq:thecondition}) is satisfied in most cases
in which $k$ is a Killing vector (up to a gauge transformation). If
it was not satisfied in any gauge, the Killing vector $k$ would be an
isometry but not a symmetry of the complete supergravity background
and would not be a generator of the vacuum supersymmetry algebra. Thus,
it must always be satisfied.
 
It is clear that these results can be generalized to higher dimensions
and supersymmetries.

\vspace{1cm}

\section*{Acknowledgments}

I would like to thank E.~Bergshoeff and specially P.~Meessen for
interesting conversations, the Institute for Theoretical Physics of
the University of Groningen for its hospitality and financial support
and M.M.~Fern\'andez for her continuous support.  This work has been
partially supported by the Spanish grant FPA2000-1584.




\begin{thebibliography}{30}


\bibitem{kn:Weyl4} H.~Weyl,
                   {\it Z.~Phys.}~330 56 (1929).
                   Translated in Ref.~\cite{kn:OR}

\bibitem{kn:OR} L.~O'Raifeartaigh,
                Princeton University Press, Princeton, New Jersey (1997).

\bibitem{kn:Kos} Y.~Kosmann,
                 {\it Annali di Mat.~Pura Appl.}~\textbf{(IV) 91}
                 (1972) 317-395.

\bibitem{Hurley:cf}
D.~J.~Hurley and M.~A.~Vandyck,
J.\ Phys.\ A {\bf 27} (1994) 4569.


\bibitem{Vandyck:ei}
M.~A.~Vandyck,
Gen.\ Rel.\ Grav.\  {\bf 20} (1988) 261.

\bibitem{Vandyck:gc}
M.~A.~Vandyck,
Gen.\ Rel.\ Grav.\  {\bf 20} (1988) 905.


\bibitem{Figueroa-O'Farrill:1999va}
J.~M.~Figueroa-O'Farrill,
Class.\ Quant.\ Grav.\  {\bf 16} (1999) 2043
[arXiv:hep-th/9902066].





\end{thebibliography}
\end{document}